\documentclass[epj,referee]{svjour}
\usepackage{textcomp}
\usepackage{epsfig}

\begin{document}
\begin{sloppy}

\title{Structural and magneto-transport characterization of 
Co$_{2}$Cr$_{x}$Fe$_{1-x}$Al Heusler alloy films}

\author{A.~D.~Rata, H.~Braak,
D.~E.~B\"{u}rgler,\thanks{\emph{Corresponding author:} 
email d.buergler@fz-juelich.de, 
phone +49 2461 614214, FAX +49 2461 614443} 
S.~Cramm, and 
C.~M.~Schneider}
%
%
\institute{Institute of Solid State Research, Electronic 
Properties (IFF6) and \\ cni -- Center of Nanoelectronic Systems for 
Information Technology, Research Center J\"ulich GmbH, D-52425 
J\"ulich, Germany}
\date{Received: date / Revised version: date}
%
\abstract{We investigate the structure and magneto-transport 
properties of thin films of the Co$_2$Cr$_{0.6}$Fe$_{0.4}$Al 
full-Heusler compound, which is predicted to be a half-metal by 
first-principles theoretical calculations.  Thin films are deposited 
by magnetron sputtering at room temperature on various substrates in 
order to tune the growth from polycrystalline on thermally oxidized 
Si substrates to highly textured and even epitaxial on MgO(001) 
substrates, respectively.  Our Heusler films are magnetically very 
soft and ferromagnetic with Curie temperatures up to 630~K. The total 
magnetic moment is reduced compared to the theoretical bulk value, but 
still comparable to values reported for films grown at elevated 
temperature. Polycrystalline Heusler films combined with MgO barriers 
are incorporated into magnetic tunnel junctions and yield 37\%
magnetoresistance at room temperature.
} 

\authorrunning{A.~D. Rata et al.}
\titlerunning{Structural and magneto-transport characterization of 
Co$_{2}$Cr$_{x}$Fe$_{1-x}$Al Heusler alloy films}
\maketitle
\thispagestyle{empty}

\section{Introduction}
The efficient injection of spin currents into non-magnetic materials, 
which is a prerequisite for many future spintronic concepts 
\cite{wolf01-1} to work, requires materials with high spin 
polarization of the conduction electrons \cite{schmidt}.  Promising 
candidates are the so-called Heusler alloys \cite{heusler}.  So far, 
many Heusler compounds are predicted to show half-metallic properties 
from first-principles calculations 
\cite{degroot,galanakis1,galanakis2}.  They are characterized by a band 
gap at the Fermi energy only for one spin direction and metallic 
properties for the other.  The charge carriers are thus 100\%
spin-polarized, rendering these materials ideal sources for efficient 
spin injection into semiconductors and for high magnetoresistance 
values in giant and tunneling magnetoresistance (GMR and TMR) devices.  
However, despite enormous experimental efforts, there has been no 
clear-cut experimental report yet on half-metallicity at room 
temperature (RT).

The Co-based full-Heusler compounds (\textit{e.g.}, 
Co$_{2}$Cr$_{x}$Fe$_{1-x}$Al, Co$_2$MnSi, and Co$_2$MnGe) have been 
intensively investigated mainly because of their high Curie 
temperatures \cite{westerholt,hutten1,kuch}.  The specific $L2_{1}$ 
structure of ordered full-Heusler alloys has the advantage of the 
absence of empty lattice sites compared to the half-Heusler compounds 
(\textit{e.g.}, NiMnSb) \cite{magprop}, which makes them less 
susceptible for site disorder.  It is widely accepted now that 
structural disorder destroys the half-metallicity.  In this respect, 
NiMnSb is a well studied example \cite{mijnarendes,ristoiu}.  Site 
disorder is thus a critical factor for deteriorating the full spin 
polarization predicted for the perfectly ordered Heusler structure, 
particularly in very thin layers used in real devices.  Yet, the 
question is to what extent site disorder is tolerable, \textit{i.e.} 
leaves at least a high spin polarization value close to 100\%.  It has 
been predicted by Miura \textit{et al.} \cite{shirai} that in the case 
of the Co$_{2}$CrAl Heusler compound in the $B2$ structural 
modification, site disorder between Cr and Al is less critical and 
allows for $84\%$ spin polarization.  Experimentally,
Block \textit{et al.} \cite{felser} have reported that 
Co$_{2}$Cr$_{0.6}$Fe$_{0.4}$Al (CCFA) bulk material in the ordered 
$L2_1$ structure exhibits 30\% magnetoresistance at RT in a
magnetic field of 1~kOe.  Shortly thereafter, Inomata \textit{et al.} 
\cite{inomata} have found 16\% TMR at RT in structures containing CCFA 
as one ferromagnetic electrode.  These findings have motivated our 
work on the Co$_{2}$CrAl Heusler compound doped with Fe.

In this study, we investigate the structure and magnetic properties of 
CCFA thin films grown by magnetron sputtering.  Various substrates 
(\textit{e.g.}, SiO$_2$, GaAs, MgO) have been employed in order to 
tune the growth from polycrystalline on thermally oxidized Si 
substrates to highly textured and even epitaxial on MgO(001) 
substrates.  Finally, we report on a high magnetoresistance value at 
RT in CCFA/MgO/CoFe junctions.

\section{Experiment}
Co$_{2}$Cr$_{x}$Fe$_{1-x}$Al thin films with various thicknesses 
are grown by magnetron sputtering on SiO$_2$ (thermally oxidized Si 
wafers), GaAs, and MgO 
substrates.  The base pressure in our chamber is below 
$1\times10^{-7}$~mbar and the sputtering Ar pressure is varied between 
$1\times10^{-3}$ and $3\times10^{-3}$~mbar.  All substrates are 
cleaned \textit{ex-situ} with solvents and immediately introduced into 
the vacuum chamber.  The MgO substrates are further cleaned by 
annealing in vacuum at 392~K, whereas no \textit{in-situ} treatment is 
done for SiO$_2$ and GaAs substrates.  We use stoichiometric 
Co$_{2}$Cr$_{x}$Fe$_{1-x}$Al targets with $x=0$ and $x=0.6$ for 
depositing our films.  The sputtering rate, measured with a quartz 
balance, is set to 0.7~\AA/s.  The temperature of the substrate during 
the sputtering process is always RT. The influence of post-growth 
annealing treatments and the use of metal (\textit{e.g.}, Cr, V) seed 
layers on the structural and magnetic properties are also 
investigated.  After deposition, the Heusler films can be annealed 
\textit{in-situ} up to 873~K. The stoichiometry is checked 
\textit{ex-situ} by secondary ions mass spectroscopy (SIMS).  We find 
that the composition of our films is similar to the target 
compositions.  The structure of the Heusler films is identified 
\textit{ex-situ} by x-ray diffraction (XRD).  Small-angle x-ray 
reflectivity (XRR) measurements are performed to determine the film 
thickness and to calibrate the sputtering rates.  The XRD measurements 
are carried out with a Philips X'Pert MRD diffractometer using Cu 
$K_\alpha$ radiation.  Conductivity measurements are performed in the 
standard four-point geometry at temperatures between 10 and 300~K. For 
the magnetic characterization, we employed the magneto-optical Kerr 
effect (MOKE) and SQUID magnetometry.  Some Heusler films are capped 
with a thin ($\approx30$~{\AA}) Cu or AlO$_x$ layer in order to 
investigate the interface properties by x-ray absorption spectroscopy 
(XAS) at the synchrotron.

\section{Results}
\subsection{Structural characterization}\label{structure}
\begin{figure}[t]
\centering{\epsfig{file=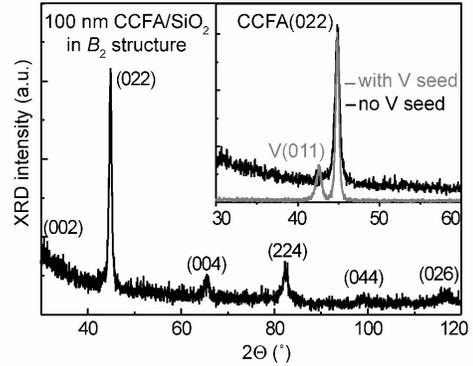,width=0.9\linewidth}}
\caption{$\theta\!-\!2\theta$ XRD measurement of a 100~nm-thick CCFA 
film deposited on SiO$_2$.  The inset compares part of the 
$\theta\!-\!2\theta$ scan around the (022) Heusler peak from films 
deposited on SiO$_2$ substrates with (grey) and without (black) a 
40~nm-thick vanadium seed layer.} \label{figure1}
\end{figure}
Figure~\ref{figure1} shows a wide $\theta\!-\!2\theta$ XRD scan from a 
100~nm-thick CCFA film grown at RT directly on a SiO$_2$ substrate.  
Despite the low deposition temperature and the absence of any seed 
layer, which normally promotes textured growth, clear diffraction 
peaks are identified in the diffraction pattern.  They indicate a 
polycrystalline morphology of the film.  Ideally, for the 
characteristic $L2_1$ structure of a Heusler compound, one should be 
able to distinguish in the XRD pattern besides the fundamental 
reflections, \textit{e.g.} (022), two types of superlattice 
reflections, \textit{i.e.} (111) and (002).  Our CCFA films grown on 
amorphous, thermally oxidized Si wafers take the so-called $B2$ (CsCl 
type) structure, where the absence of the (111) peak indicates 
complete disorder between Al and (Cr, Fe) atoms, while Co atoms occupy 
the correct sublattice.  The same type of disordered structure has 
previously been observed in CCFA films by Inomata \textit{et al.} 
\cite{inomata}.  When GaAs is used as a substrate (not shown), 
polycrystalline films with the same $B2$ structure could only be 
obtained when a seed layer, \textit{e.g.} Cr or V, is deposited prior 
the Heusler film.  Otherwise we cannot observe any crystalline peaks 
in the XRD spectra, except for a broad (022) reflection on top of an 
amorphous background.

\begin{figure}[b]
\centering{\epsfig{file=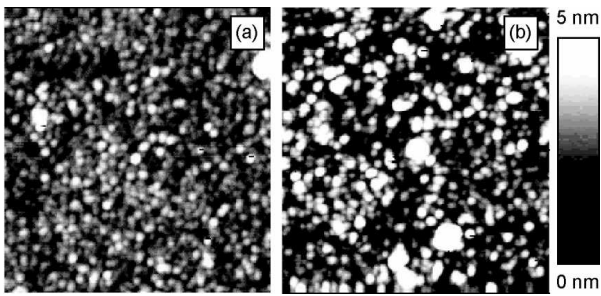,width=0.9\linewidth}}
\caption{Representative AFM pictures taken from 100~nm-thick CCFA 
films deposited on SiO$_2$ without (a) and with (b) a vanadium seed 
layer.  The size of the scans is $1\times1$~$\mu$m$^{2}$.}
\label{figure2}
\end{figure}
It has been shown by Geiersbach \textit{et al.} \cite{westerholt} 
that, in order to grow highly textured Co-based Heusler thin films, 
vanadium is the best choice as a seed layer among Al, Cr, Cu, and V. 
In the inset of Fig.~\ref{figure1} we compare $\theta\!-\!2\theta$ 
scans around the (022) diffraction peak from a 100~nm-thick CCFA film 
grown on SiO$_2$ substrates with and without a 40~nm-thick vanadium 
seed layer, respectively.  We also observe that the vanadium seed 
layer improves the texture of the Heusler films and that the 
amorphous-like background is completely suppressed.  Further 
structural analysis is done by atomic force microscopy (AFM).  In 
Fig.~\ref{figure2}, we display two representative AFM pictures taken 
on the CCFA films discussed in the inset of Fig.~\ref{figure1}.  Quite 
surprising, the CCFA film grown directly on the SiO$_2$ substrate is 
very smooth with a rms roughness of only 0.5~nm.  A uniform 
distribution of grains with an average diameter of 40~nm is observed 
in Fig.~\ref{figure2}(a).  On the contrary, when vanadium is used as a 
seed layer, the grains' size is not as uniform and the roughness 
increases to $\approx$ 1.0~nm.  The strain induced by the large 
lattice mismatch between vanadium and the CCFA Heusler film might be 
an explanation for these observations.

Temperature is an important parameter in optimizing the structural and 
magnetic properties of Heusler compounds.  Co-based alloy films, as 
for example Co$_2$MnSi, Co$_2$MnGe, and Co$_2$MnSn, with ordered 
crystallographic structure could only be obtained by growing 
\cite{westerholt} or annealing \cite{hutten1} at relatively high 
temperatures above 700~K. Our CCFA films grow with good crystalline 
order already at RT. In order to study the influence of 
temperature on the structure and magnetic properties, we perform 
\textit{in-situ} post-growth annealing experiments.  The 
\textit{in-situ} annealing treatment at temperatures up to 773~K has 
only little influence on the crystallographic structure of our films.  
We only observe a slight increase in the intensity of the diffraction 
peaks.  On the other hand, changes in the magnetic properties are 
observed and will be discussed in Section~\ref{magnetic}.

\begin{figure}[t]
\centering{\epsfig{file=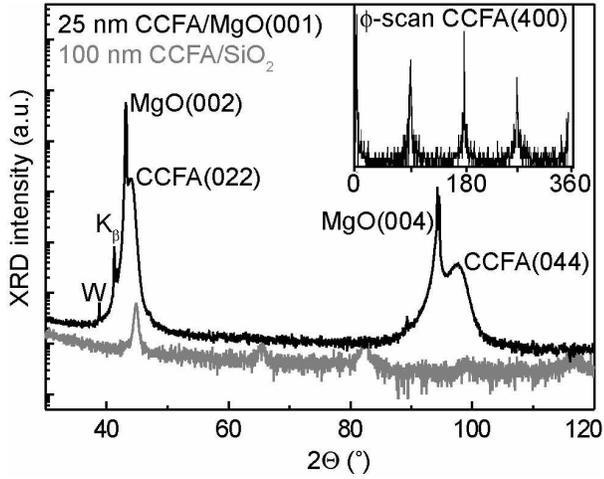,width=0.9\linewidth}}
\caption{$\theta\!-\!2\theta$ XRD measurement of a 25~nm-thick CCFA 
film deposited on MgO(001) (black) in comparison to the data of 
Fig.~\ref{figure1} (grey).  Only the (022) and (044) diffraction peaks 
can be observed close to the (002) and (004) peaks characteristic of 
the MgO(001) substrate.  MgO(002) Cu \textit{K$_{\beta}$} and W 
\textit{L}$_{\alpha}$ peaks are indicated on the left of the (002) 
main MgO peak.  Inset: In-plane $\phi$-scan through the (400) peaks of 
CCFA.} \label{figure3}
\end{figure}
It is well known that the choice of the substrate can drastically 
influence the structural characteristics and physical properties of a 
film.  In order to improve the structure of the CCFA Heusler films we 
have chosen a single crystalline substrate, namely magnesium oxide.  
Kelekar and Clemens \cite{clemens1,clemens2} prepared epitaxial CCFA 
films by magnetron sputtering onto a MgO(001) substrate held at 773~K. 
In Fig.~\ref{figure3} we present our XRD results of a 25~nm-thick CCFA 
film grown at RT on MgO(001) (black) in comparison with the data of 
the CCFA/SiO$_{2}$ system from Fig.~\ref{figure1} (note the 
logarithmic scale in Fig.~\ref{figure3}).  The data clearly indicate 
epitaxial growth of CCFA films on MgO(001) at RT. Only the (022) and 
(044) diffraction peaks of the film are observed close to (002) and 
(004) peaks of the MgO(001) substrate, indicating a highly oriented 
CCFA film on top of the rocksalt MgO substrate.  The peaks originating 
from the film are broadened due to its finite thickness (25~nm).  It 
is interesting to note that the [011] direction of the CCFA films is 
parallel to [001] of the MgO substrate.  Thus, our films take a 
different orientation with respect to the MgO substrate compared to 
the results reported in Refs.~\cite{clemens1,clemens2}, where 
CCFA[001]//MgO[001].  The lattice constant normal to the surface can 
be calculated from $\theta\!-\!2\theta$ scans by using Bragg's law.  
We obtain a value of 5.81~{\AA}, which is very close to the bulk value 
of 5.88~{\AA} \cite{felser}.  The lattice constant of MgO is 
4.21~{\AA}, so the relatively small lattice mismatch (-1.3\%) after a
45{\textdegree} in-plane rotation makes the MgO(001) surface a good 
template for epitaxial growth of CCFA films.  For our epitaxial 
relationship, the unit cell of CCFA in the (011) plane at the 
interface is rectangular with an aspect ratio of $\sqrt{2}$.  After 
the 45{\textdegree} rotation, the short edge points along either 
MgO[110] or MgO[-110] directions giving rise to two possible 
structural domains.  Polar in-plane ($\phi$-)scans in the (011) plane 
of CCFA obtained by tilting the sample normal out of the scattering 
plane, reveal four-fold symmetry and, therefore, give definitive 
evidence for the epitaxial growth in two structural domains at RT. An 
example of a polar scan through the (400) Heusler peaks is shown in 
the inset of Fig.~\ref{figure3}.  Small angle XRR measurements show 
smooth and well defined interfaces between the Heusler film and the 
MgO substrate.  A representative XRR spectrum is depicted in 
Fig.~\ref{figure4}.  
\begin{figure}[t]
\centering{\epsfig{file=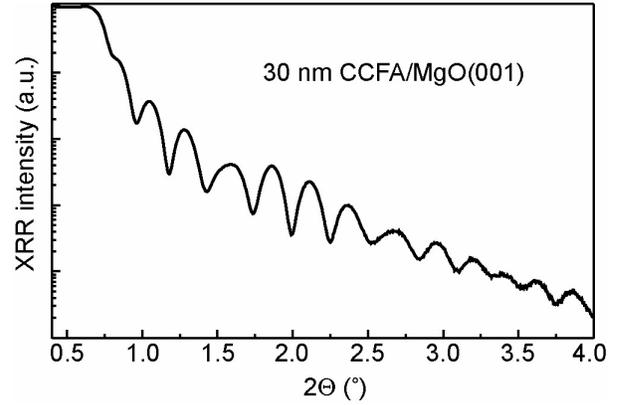,width=0.9\linewidth}}
\caption{Small angle XRR scan showing a well defined interface between 
the CCFA(011) film and the MgO(001) substrate.}
\label{figure4}
\end{figure}

\subsection{Magnetic characterization}\label{magnetic}
\begin{figure}[t]
\centering{\epsfig{file=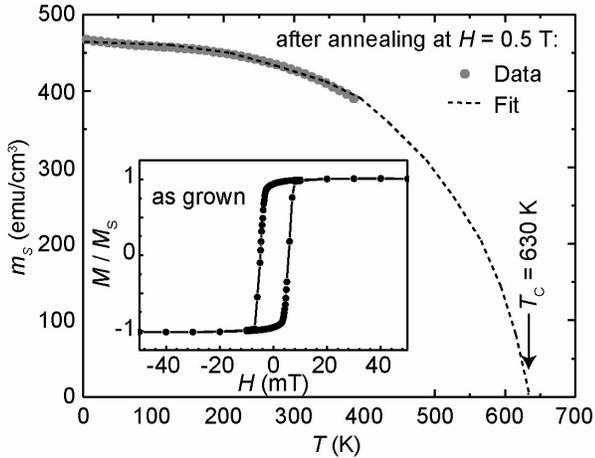,width=0.9\linewidth}}
\caption{Inset: Normalized magnetic hysteresis curve measured by SQUID 
at RT of an as-grown 100~nm-thick CCFA film on SiO$_2$.  The main graph 
shows the temperature dependence of the saturation magnetization 
$m_{S}$ of the same film after annealing in vacuum at 773~K for 1~hour 
together with the fitted Brillouin function.  The Curie temperature is 
about 630~K.} \label{figure5}
\end{figure}
Our CCFA Heusler films display a soft ferromagnetic behavior.  In the 
inset of Fig.~\ref{figure5} we present a normalized $M$--$H$ loop of a 
100~nm-thick CCFA film grown on a SiO$_2$ substrate.  The coercive 
field is approximately 5~mT. Our films are magnetically softer 
compared to the recently reported results about CCFA films deposited 
on Al$_2$O$_3$ substrates, which had a coercive field of 10~mT 
\cite{jacob}.  The Curie temperature of an as-grown film on a SiO$_2$ 
substrate is estimated to be about 540~K from the extrapolation of the 
magnetization \textit{versus} temperature curve (not shown).  
Annealing in vacuum at 773~K for one hour significantly increases the 
Curie temperature as shown by the temperature dependence of the 
saturation magnetization after annealing in Fig.~\ref{figure5}, where 
the solid line is a fitted Brillouin function yielding $T_{C}\approx 
630$~K.

\begin{table}[b]
\caption{Magnetic moment $m_{s}$ at 5~K, Curie temperature $T_{C}$, and 
coercive field $H_{C}$ of CCFA films grown on different 
substrates in the as-grown and 773~K post-annealed state.}
\label{table}      
\begin{tabular}{lcc|ccc}
\hline\noalign{\smallskip}
Substrate & thickness & state & $m_{S}$    & $T_{C}$ & $H_{C}$ \\ 
          & (nm)      &       & ($\mu_{B}$/f.u.)   &  (K)    & (mT)    \\ 
\noalign{\smallskip}\hline\noalign{\smallskip}
SiO$_2$ & 100 & as-grown & 2.42 & 540 & 5 \\ 
        & 100 & annealed & 2.56 & 630 & 5-6 \\ \hline 
40~nm V & 100 & as-grown & 2.43 & 575 & 7 \\ 
        & 100 & annealed & 2.52 & 590 & 7 \\ \hline 
MgO     &  25 & as-grown & 2.46 & 380 & 6 \\ 
        &  25 & annealed & 2.56 & 500 & 6 \\ \hline 
\noalign{\smallskip}\hline
\end{tabular}
\end{table}
The total magnetic moment obtained from the saturation magnetization 
at 5~K is 442~emu/cm$^{3}$ corresponding\footnote{The volume per f.u.  
is (5.88~{\AA})$^{3}$/4 and 1~emu~= 1.079$\times10^{20}$~$\mu_{B}$.} 
to 2.42~$\mu_{B}$ per formula unit (f.u.)  for as-grown films on 
SiO$_2$.  After annealing in vacuum this value increases to 
2.56~$\mu_{B}$/f.u.  (467~emu/cm$^{3}$).  Magnetic moments, Curie 
temperatures, and coercive fields for CCFA films grown on various 
substrate systems are compiled in Table~\ref{table}.  The lower Curie 
temperatures for the film grown on MgO might be related to the reduced 
thickness.  The magnetic properties do not strongly vary for different 
substrates, in spite of the quite different film structures.  But in 
all cases the post-annealing increases the magnetic moment as well as 
the Curie temperature, which both are governed by the film volume.  
All films are relatively soft with $H_{C}$ in the range from 5 to 
7~mT. These findings are in agreement with results from FMR 
measurements of our CCFA samples performed by Rameev \textit{et al.} 
\cite{rameev}.  
\begin{figure}[t]
\centering{\epsfig{file=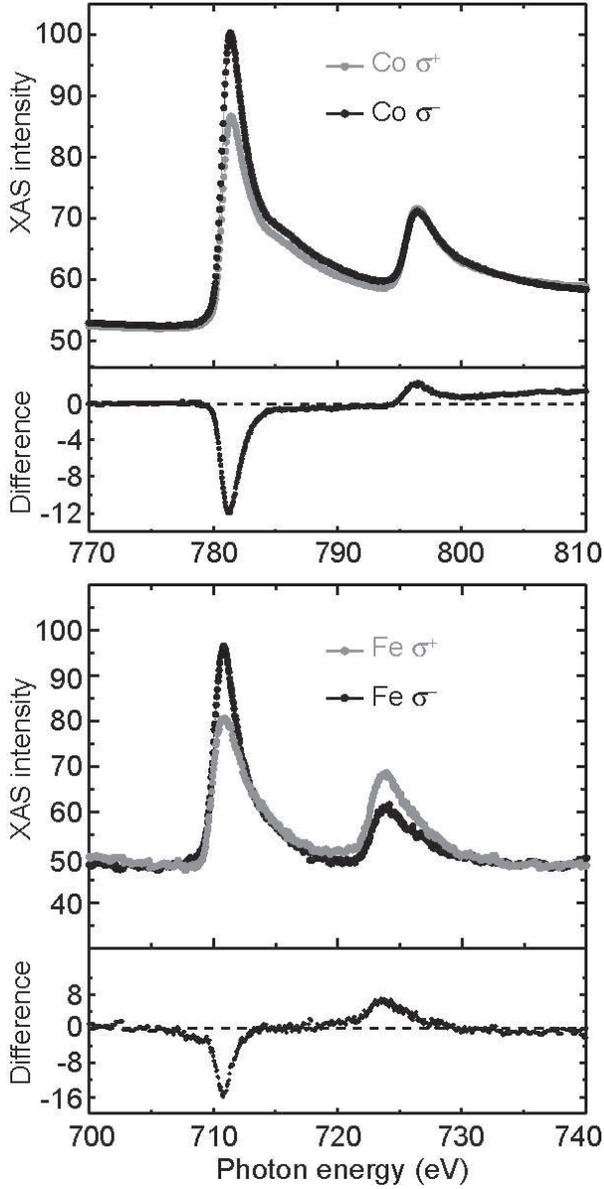,width=0.9\linewidth}}
\caption{XMCD-XAS spectra at the Co and Fe $L_{23}$ edges show clear 
magnetic signals and indicate parallel alignment of the Co and Fe 
moments.} \label{figure6}
\end{figure}

The measured magnetic moments per formula unit are small compared to 
the theoretical value for ordered CCFA with $L2_1$ structure, which is 
3.8~$\mu_{B}$/f.u.  \cite{galanakis2}.  On the other hand, our values 
are comparable to the data reported for films grown at elevated 
temperatures \cite{clemens1,clemens2}.  These authors also found a 
reduced experimental magnetic moment as compared to the calculated 
moment for Cr-containing samples.  Only the magnetic moment of 
Co$_2$FeAl reaches the predicted value of 4.9~$\mu_{B}$/f.u.  
\cite{galanakis3}.  Moreover, even the magnetic moment measured in 
bulk samples is smaller than the calculated one 
\cite{elmers1,elmers2}.  It is generally recognized that the magnetic 
moments and the degree of spin polarization are critically dependent 
on the chemical disorder \cite{picozzi}.  Despite the fact that our 
CCFA Heusler films have a rather good crystallographic order, site 
disorder seems to play a critical role for the Curie temperature and 
the total magnetic moment.  It is not possible to determine the site 
disorder in CCFA films from XRD, because Co, Cr, and Fe have very 
similar scattering factors.  Neutron scattering experiments could give 
more insight about the degree of chemical disorder in the Heusler 
films.

We gain additional information on the presence of site disorder in our 
films from XAS by measuring the x-ray magnetic circular dichroism 
(XMCD).  CCFA films, covered with a thin AlO$_x$ cap layer are 
investigated exciting with circular polarized light at the Co, Fe, and 
Cr $L_{23}$ edges.  The measurements are done in the total-yield mode, 
where one directly measures the sample current while scanning the 
photon energy.  The finite escape depth of the photoelectrons limits 
the information depth to the interface region.  The sample current is 
normalized to the photon intensity measured on a gold mesh.  The 
magnetic field applied to the sample (0.25~T) is aligned with the 
surface normal and lies at an angle of about 30{\textdegree} with 
respect to the incident photon direction.  The line shape of the XAS 
spectra is that of metallic (\textit{e.g.} non-oxidized) Co and Fe.  
We obtain a clear dichroic signal at both the Co and Fe edges, which 
indicates for the interface region ordered Co and Fe spins with 
parallel alignment (see Fig.~\ref{figure6}).  In contrast, no magnetic 
dichroism is observed at the Cr edge (not shown).  This is an 
indication for partially antiparallel alignment of the Cr interfacial 
spins with respect to each other, which can be related to the presence 
of site disorder.  This result could also explain the reduced values 
of the Curie temperature and the magnetic moment compared to the bulk 
material.  Moreover, we find that a selective oxidation of Cr always 
occurs for CCFA capped with 3~nm of Cu, while 3~nm-thick AlO$_x$ 
layers protect the CCFA films against oxidation.  A quantitative 
analysis of XAS-XMCD measurements performed on CCFA Heusler compounds 
has been reported by Elmers \textit{et al.} \cite{elmers1,elmers2}.

\subsection{Transport properties}\label{transport}
\begin{figure}[t]
\centering{\epsfig{file=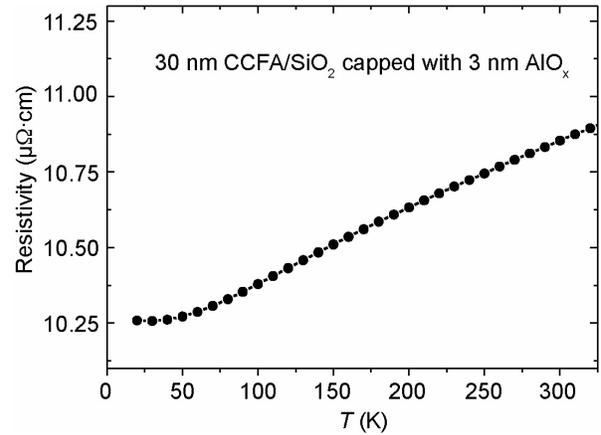,width=0.9\linewidth}}
\caption{In-plane resistivity of a 50~nm-thick CCFA film covered
with a 3~nm-thick AlO$_x$ cap layer indicating metallic conductivity
down to 10~K.} \label{figure7}
\end{figure}

Resistivity measurements with the current flowing in the plane of the 
sample reveal a metallic behavior, with the resistivity increasing 
with increasing temperature as shown in Fig.~\ref{figure7} for a 
50~nm-thick CCFA film covered with 3~nm of AlO$_x$.  The residual 
resistivity at 10~K is quite low, $\approx10^{-5}~\Omega$cm, and 
the temperature dependence is rather weak.  This anomalous behavior 
of the temperature-dependent resistivity has also been observed in 
Ni-based Heusler compounds, and it was interpreted as a 
characteristics of disordered metallic systems \cite{maydos}.

Recently, relatively high TMR ratios have been obtained using Co-based 
full-Heusler alloy thin films, \textit{i.e.} Co$_2$MnSi 
\cite{hutten2}, Co$_2$FeAl \cite{okamura}, and CCFA 
\cite{inomata,inomata1,marukame}.  Owing to the good structural quality 
and the well oriented growth of our CCFA films on MgO, we have 
integrated them into TMR structures with MgO as tunneling barrier.  
MgO has recently been proven to be a well behaved barrier material 
yielding very high TMR ratios in fully epitaxial Fe/MgO/Fe 
\cite{yuasa} and sputtered CoFe/MgO/CoFe structures with highly 
oriented MgO barriers \cite{parkin}.

\begin{figure}[b]
\centering{\epsfig{file=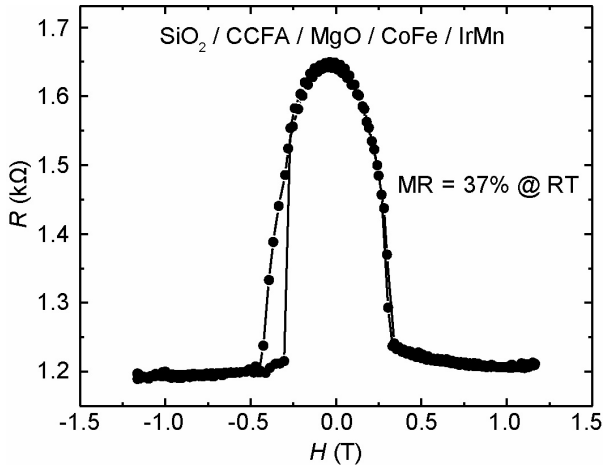,width=0.9\linewidth}}
\caption{Magnetoresistance curve of a $10\times 10$~$\mu$m$^{2}$ 
junction with layer sequence 
SiO$_2$/CCFA(25~nm)/MgO(3~nm)/CoFe(5~nm)/IrMn(15~nm).  The measurement 
is performed at RT and yields a magnetoresistance of 37\%.}
\label{figure8}
\end{figure}
The layered TMR structures are prepared by magnetron sputtering 
without breaking the vacuum with the following layer sequence: 
SiO$_2$/ CCFA(25~nm)/MgO(3~nm)/CoFe(5~nm)/IrMn(15~nm).  In some cases 
a 40~nm-thick MgO layer is deposited on the substrate to improve the 
texture of the CCFA electrode.  All layers in the structure are 
successively sputtered at RT. The MgO barrier is deposited by 
rf-stimulated discharge from a stoichiometric target.  The sputtering 
rate is calibrated by measuring the thickness of a MgO film by XRR. 
Here we present results for magnetic tunnel junctions containing CCFA 
as bottom ferromagnetic electrode, deposited directly on a SiO$_2$ 
substrate.  The completed stack is annealed \textit{in-situ} for 
1~hour at 523~K. Although an IrMn antiferromagnet is employed to pin 
the magnetization of the upper CoFe electrode, the two ferromagnetic 
layers do not switch separately, as observed in SQUID measurements.  
We mention at this point that no magnetic field can be applied during 
the \textit{in-situ} annealing.  Thus, no magnetic alignment by field 
cooling is introduced to this structure to induce an exchange bias.  
Junctions with an area from $3\times3$ up to $15\times15$~$\mu$m$^{2}$ 
with crossed electrodes are patterned for magnetotransport 
measurements in the current-perpendicular-plane (CPP) geometry by 
standard optical lithography.

In Fig.~\ref{figure8} we show the magnetoresistance curve measured at 
RT. The TMR ratio defined as TMR = $(R_{AP}- R_P)/R_P$, where $R_{AP}$ 
and $R_P$ are the resistances for the antiparallel and parallel 
magnetization configurations, respectively, reaches 37\%. The 
resistance-area product is about 100~k$\Omega\mu$m$^{2}$.  This is a
relatively large value for structures comprising a ferromagnetic 
Heusler electrode.  Further investigations of the magnetotransport 
properties of TMR structures grown with the optimized parameters reported 
in this work are in progress and will be detailed in an upcoming 
report \cite{rata_tb}.

\section{Summary}
To summarize, we succeeded in preparing high quality 
Co$_2$Cr$_{0.6}$Fe$_{0.4}$Al (CCFA) Heusler alloy thin films using 
magnetron sputtering at RT. Various substrates (SiO$_2$, GaAs, MgO) 
were employed in order to tune the growth from polycrystalline on 
thermally oxidized Si substrates to highly textured and even epitaxial 
on MgO(001) substrates.  We studied the influence of post-growth 
\textit{in-situ} annealing treatments and the use of Cr and in 
particular V seed layers on the structural and magnetic properties.  
Magnetization measurements revealed ferromagnetic ordering.  The Curie 
temperatures could be increased up to 630~K after annealing in vacuum 
at 773~K. XRD measurements indicated a high quality of the films grown 
on MgO substrates with well defined interfaces and low roughness, 
while the films grow in a polycrystalline fashion on SiO$_2$.  
Although our films grow at RT in a different orientation on the 
MgO(001) substrate than reported in Refs.~\cite{clemens1,clemens2}, 
the total magnetic moments are very similar with values of about 
2.5~$\mu_{B}$/f.u.  Therefore, the structural order is less important 
for the size of the magnetic moment.  This is also reflected by the 
rather constant value of the magnetic moment for the various 
substrates, which give rise to quite different structural properties.  
In accordance with the theoretical result of Picozzi \textit{et al.} 
\cite{picozzi} we conclude that the site disorder is the relevant 
parameter which determines the size of the magnetic moment and the 
Curie temperature in thin films.

Finally, we observed large magnetoresistance values of up to $\approx 
40$\% at RT
in TMR structures containing a polycrystalline CCFA film directly 
grown on SiO$_2$ as a bottom ferromagnetic electrode and MgO as a 
tunneling barrier, both deposited by magnetron sputtering at RT. These 
preliminary transport measurements are encouraging for further 
investigations of TMR structures consisting of highly-oriented or 
epitaxial Co-based full-Heusler alloy films combined with possibly 
epitaxial MgO tunneling barriers.  We expect a strong dependence on 
the film structure (\textit{i.e.} choice of the substrate), much 
higher TMR ratios, and insight into the half-metallic properties of 
Heusler alloys at the interface, because the tunneling effect is very 
sensitive to the interface properties.

\begin{acknowledgement}

We would like to thank F.-J.~K\"{o}hne for technical support, 
P.~Erhart for the possibility of using the x-ray diffractometer, 
H.~Elmers, M. Le\v{z}ai\'{c}, and P.~Gr\"{u}nberg for discussions and 
useful suggestions.

\end{acknowledgement}

\end{sloppy}
\end{document}